\documentclass[aps,prl,twocolumn,showpacs,amsfonts]{revtex4-1}

\usepackage{amsmath, amssymb}
\usepackage{amsfonts}
\usepackage{graphicx}
\usepackage{bm}
\usepackage[usenames,dvipsnames]{color}

\newcommand{\pd}[2]{\frac{\partial #1}{\partial #2}}

\begin{document}
\title{A simple nonlinear equation for structural relaxation in glasses}

\author{Itamar Kolvin$^1$ and Eran Bouchbinder$^2$}
\affiliation{$^1$ Racah Institute of Physics, Hebrew University of
Jerusalem, Jerusalem 91904, Israel\\
$^2$ Chemical Physics Department, Weizmann Institute of Science,
Rehovot 76100, Israel}

\begin{abstract}
A wide range of glassy and disordered materials exhibit complex,
non-exponential, structural relaxation (aging). We propose a simple nonlinear rate equation $\dot{\delta}\!=\!
a\left[1\!-\!\exp\left(b\,\delta\right)\right]$, where $\delta$ is
the normalized deviation of a macroscopic variable from its
equilibrium value, to describe glassy relaxation. Analysis of
extensive experimental data shows that this equation quantitatively
captures structural relaxation, where $a$ and $b$ are both temperature-, and more importantly, history-dependent
parameters. This analysis explicitly demonstrates that structural relaxation cannot be accurately described by a single non-equilibrium variable.
Relaxation rates extracted from the data imply the existence
of cooperative rearrangements on a super-molecular scale.
\end{abstract}


\maketitle

Glass-forming materials exhibit rapid increase in relaxation
timescales when going through their glass temperature $T_g$
\cite{AngellReview}. When external conditions change abruptly, the
observation of the full relaxation to equilibrium becomes
exceedingly difficult, except for a narrow range of
temperatures near $T_g$. The relaxation is characteristically
non-exponential and spans several orders of magnitude in time. In
many cases, the relaxation is logarithmic. This behavior is
paralleled in a broad range of disordered systems: compaction of
granular materials \cite{Nagel_angle,Nagel_density}, crumpling of
thin sheets \cite{Nagel_crumpling,Mokhtar}, aging of contact area in
dry friction \cite{Jay}, aging of conductivity in electron glasses
\cite{Ovadyahu, Amir} and mechanical relaxation of star polymers in
gels \cite{deGennes,star_polymer}. This apparently wide-spread
behavior might suggest a generic origin of slow glassy relaxations (aging).

Glassy relaxation is typically probed by tracking the time evolution
of a macroscopic quantity, e.g. the volume or the enthalpy of a
sample, in response to an abrupt change in an externally controlled
variable, e.g. the temperature. In the latter case, when an initial
temperature $T_0$ is rapidly changed to $T$, in the vicinity of the
glass temperature $T_g$, various degrees of freedom of the glass
respond differently. The vibrational degrees of freedom quickly
equilibrate at the new temperature $T$. The structural degrees of
freedom, however, carry long-time ``memory'' of the original state
at $T_0$ and fall out-of-equilibrium with $T$. It is the
out-of-equilibrium dynamics of the structural degrees of freedom
towards a new equilibrium at $T$ that is at the heart of
``structural relaxation''.

Structural relaxation is conventionally interpreted in terms of the
Tool-Narayanaswamy-Moynihan (TNM) \cite{Moynihan}, or its equivalent
KAHR \cite{KAHR}, phenomenological 4-parameter models. The main assumption in
these models is that during relaxation, a single dynamical variable is
sufficient for describing the non-equilibrium state of the glass.
In the case of the TNM model this variable is a ``fictive temperature'', defined to be a linear function
of the probed property, e.g. volume.

However, it is known that these models do not describe experimental
data accurately \cite{AngellReview}, and actually fail to account for thermal history
dependence \cite{Simon2006}. In addition, while it has been recognized that non-monotonic relaxation (i.e. the Kovacs memory effect \cite{Kovacs1963}) cannot be described by a single non-equilibrium state variable \cite{MS-04, LEUZZI-09}, some recent works have suggested that this might be possible for monotonic relaxations of the type considered here \cite{MS-04, BLII-09, BL-Kovacs-10}.

In spite of the seemingly universal nature of structural relaxation
in glasses, as well as its great scientific and technological
importance, a theoretical understanding of it is still missing.
In this Letter, we propose a simple, analytically solvable, nonlinear
rate equation for describing structural relaxation (aging) in glasses. The
proposed equation is shown to quantitatively capture extensive
experimental measurements on volume relaxation, yet it explicitly demonstrates the inadequacy of
a single non-equilibrium description of glassy relaxation. This analysis allows us to shed light on some basic properties of structural relaxation in
these systems, including estimates of activation energy barriers and volumes.

Volume relaxation of glassy materials is usually studied using
mercury dilatometry \cite{PolymerChemistry}. The classical experiments
in volume relaxation were performed by Kovacs more than five decades ago \cite{Kovacs1963}. These
measurements, which in extreme cases reached months,
are routinely used and provide a
standard testing ground for models. In these so-called down-jump
experiments, a glass sample is rapidly
quenched from equilibrium at $T_0$ to a lower temperature $T$.
Measurements begin at $t_i$, the characteristic time it takes
for the vibrational degrees of freedom to thermalize.

\begin{figure}[!ht]
\includegraphics[scale=0.5]{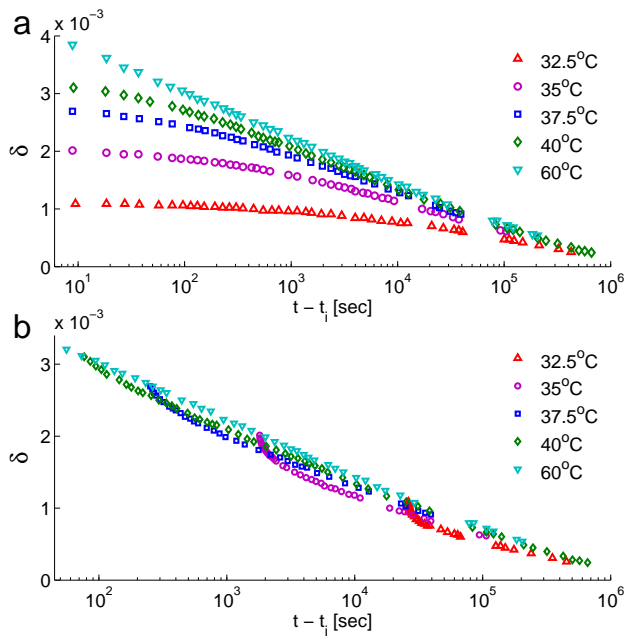}
\caption{(color online) (a) Down-jumps with a fixed target temperature $T\!=\! 30^oC$ in PVAc.
Data digitized from Fig. 16 in \cite{Kovacs1963}. (b) The above measurements time-shifted
onto the $60^oC\rightarrow 30^oC$ curve.}
 \label{Fig1}
\end{figure}

The first question we raise is whether structural relaxation can be properly described by a single non-equilibrium variable model.
To address this question, we denote such a variable by $\delta(t)$ and note that a single variable description means that the rate
of relaxation is uniquely determined by the instantaneous value of the variable, i.e. $\dot\delta(t)\!=\!r[\delta(t)]$, where $r[\cdot]$ is some functional. Therefore, in the framework of such models
there exists a function $g(t)$ such that $\delta(t) \!=\! g(t+g^{-1}(\delta_0))$,
where $\delta_0\!=\!\delta(0)$ is the initial condition, and hence a set of measurements which differ \textit{only} in
initial condition could be time-shifted in such a way
that all curves would collapse on a single master curve.
An example of such a set is shown in Fig. \ref{Fig1}(a)
with measurements digitized from Kovacs' original work \cite{Kovacs1963},
given in terms of $\delta \!=\! (V-V_\infty)/V_\infty$,
the normalized deviation of the volume from its asymptotically
stable value $V_\infty$. Here the temperature $T \!=\! 30^oC$ is the \textit{same} for all measurements,
and only the initial state of the system is changed by quenching from
different initial temperatures $T_0$. In Fig. \ref{Fig1}(b), we
time-shift each one of the curves such that their initial values
would sit on the $60^oC\rightarrow 30^oC$ curve. The failure of the time-shifted curves to collapse on a master curve implies that using a single
non-equilibrium variable (here the deviation of the volume from its equilibrium value) would be
inadequate for constructing a predictive model of structural relaxation.

This observation seems to be in line with the common knowledge that glassy relaxation is characterized by a broad spectrum of relaxation
times. A fundamental modeling approach would incorporate these
various timescales in the time evolution of the probability
distribution function of the volume $v$ of mesoscopic material
elements $f(v,t)$, as it approaches the stationary equilibrium
distribution $f_T^{eq}(v)$ at $T$ during structural relaxation. This
is a daunting task that has been pursued only in simple models
\cite{SOLLICH-97, 03BBDG}. Our goal here is to show that while
models that use only the macroscopic volume
$V(t)\!=\!N\int_0^\infty\!v f(v,t)dv$ (where $N$ is the number of
elements) cannot be complete, they can still teach us something and might serve as
a starting point for constructing an adequate phenomenological model.

\begin{figure*}[!ht]
\includegraphics[scale=0.55]{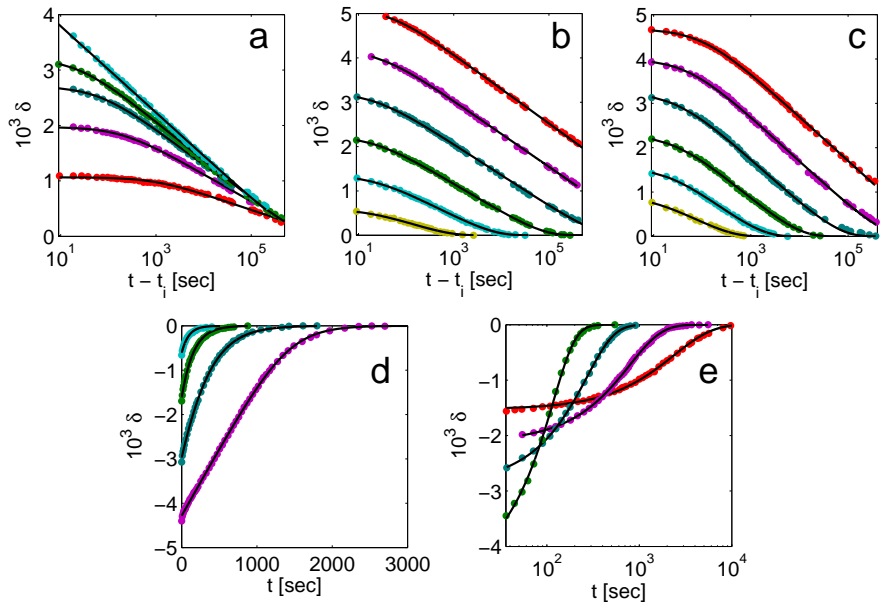}
\caption{(color online) Fits of Eq. (\ref{sol}) to Kovacs' data
\cite{Kovacs1963}. The solid black curves are three-parameter fits.
(a) Down-jumps in PVAc with fixed $T\!=\!30^\circ C$. See Fig. \ref{Fig1}.
(b) Down-jumps in PVAc from the same initial condition $T_0\!=\!40^\circ C$
to $T \!=\!25,...,37.5^\circ C$ (top to bottom).
(c) Down-jumps in glucose from a fixed initial temperature $T_0\!=\!40^\circ C$
 to $T \!=\! 19.8,...,32.5^\circ C$ (top to bottom).
(d) Up-jumps in PVAc. Note the linear scale.
Here $T_0 \!=\! 30,..,37.5^\circ C$ (bottom to top) to  $T \!=\! 40^\circ C$.
The measurement starts at $t_i \!=\! 36$ sec.
(e) Up-jumps in glucose where $T_0
\!=\! 25^\circ C$ is fixed while $T \!=\!  30,..,37.4^\circ C$ (bottom to top). The
measurement starts at $t_i \!=\! 36$ sec.} \label{Fig2}
\end{figure*}

There are two basic approaches for understanding logarithmic
relaxations. The first approach sees the relaxation as a linear
response, i.e. relaxation rates are independent of the state of
the system. A logarithmic response can then be obtained by
summing over a spectrum of exponential relaxation modes.
This approach was suggested in \cite{Kovacs1963}, pursued in
\cite{KimmelUhlmann}, and was recently also invoked in the context of
 electron glasses \cite{Amir}. Essentially, the evolution of the deviation from
equilibrium is assumed to take the form $\delta(t)\! \propto
\!\int_0^\infty\!p(\lambda)\exp(-\lambda\,t)d\lambda$, where
$p(\lambda)$ is the distribution of relaxation rates. When
$p(\lambda)\!\sim\!\lambda^{-1}$ in a certain range, logarithmic
behavior emerges. The second approach, suggested in many contexts
(e.g. in \cite{Nagel_crumpling}), describes logarithmic relaxation
as a result of the dependence of rate on instantaneous state,
e.g. slowing-down due to compaction. A nonlinear equation of the form
$\dot\delta\!\propto\! - \exp(b\,\delta)$ is then proposed to yield
$\delta\!\propto\!-\log(t)/b$.

We follow the latter approach and propose to describe structural relaxation by the following equation
\begin{equation}
\dot{\delta} = a \left(1- e^{b\,\delta}\right)\ . \label{model}
\end{equation}
Here $a$ is a basic relaxation rate and $b$ is a constant. The $1$
added in brackets ensures that $\dot{\delta}$ vanishes with $\delta$.
Equation (\ref{model}) admits the analytic solution
\begin{equation}
\delta(t) = - \frac{1}{b}\log\left[1-(1-e^{-b\,\delta_0})\,e^{-a b
\,t}\right]\,,\label{sol}
\end{equation}
where $\delta_0$ is the initial condition. For large enough initial
amplitudes, for which $\exp(-b\,\delta_0) \!\ll\! 1$, and short
times $a b \,t \ll 1$, we have
$\delta(t)\!\simeq\!-\log(a b\,t)/b$. The final stage of
the relaxation is exponential, $\delta(t)\!\propto\!\exp(-a b\,t)$.
For large negative initial amplitudes, for which $\exp(-b\,\delta_0)
\!\gg\! 1$, and short times $a\,t \!\ll\!-\delta_0$, the relaxation
is linear $ \delta \simeq \delta_0 + a\,t$. Again, the final
stage of relaxation is exponential, $\delta(t)\!\propto\!-\exp(-a
b\,t)$. This marked asymmetry between $\delta>0$ and $\delta<0$ relaxations naturally
emerges from the exponential $\delta\!\to\!-\delta$ asymmetry in Eq. (\ref{model}).

To test the model we present in Fig. \ref{Fig2} fits of Eq. (\ref{sol}) to a significant portion of Kovacs' original data \cite{Kovacs1963}.
For each curve the parameters ${\delta_0,a,b}$ were independently
varied. Note, however, that $\delta_0$ is essentially determined by the first data point and hence is not a real fitting parameter. Down-jumps in PVAc with fixed target temperature $T$ (panel (a))
and fixed initial condition $T_0$ (panel (b)) are satisfactorily captured
by Eq. (\ref{sol}). We checked the equation also against data for
glucose (panel (c)) which appears in \cite{Kovacs1963}. Panels (d-e)
show fits to up-jumps - experiments where the initial temperature $T_0$
is lower than $T$, making the initial condition $\delta_0 < 0$.
Panel (d) is plotted in a \textit{linear} scale to show the manifestly
linear portion of the relaxation. Usually both up-jumps and down-jumps
are plotted using a logarithmic time scale to show the asymmetry
between the two responses. Here we suggest that this asymmetry can be
understood to emerge from the nonlinearity of the response.
Additional data were shown to be consistent with Eq. (\ref{model}) \cite{sources}.

\begin{figure}[ht]
\includegraphics[scale=0.56]{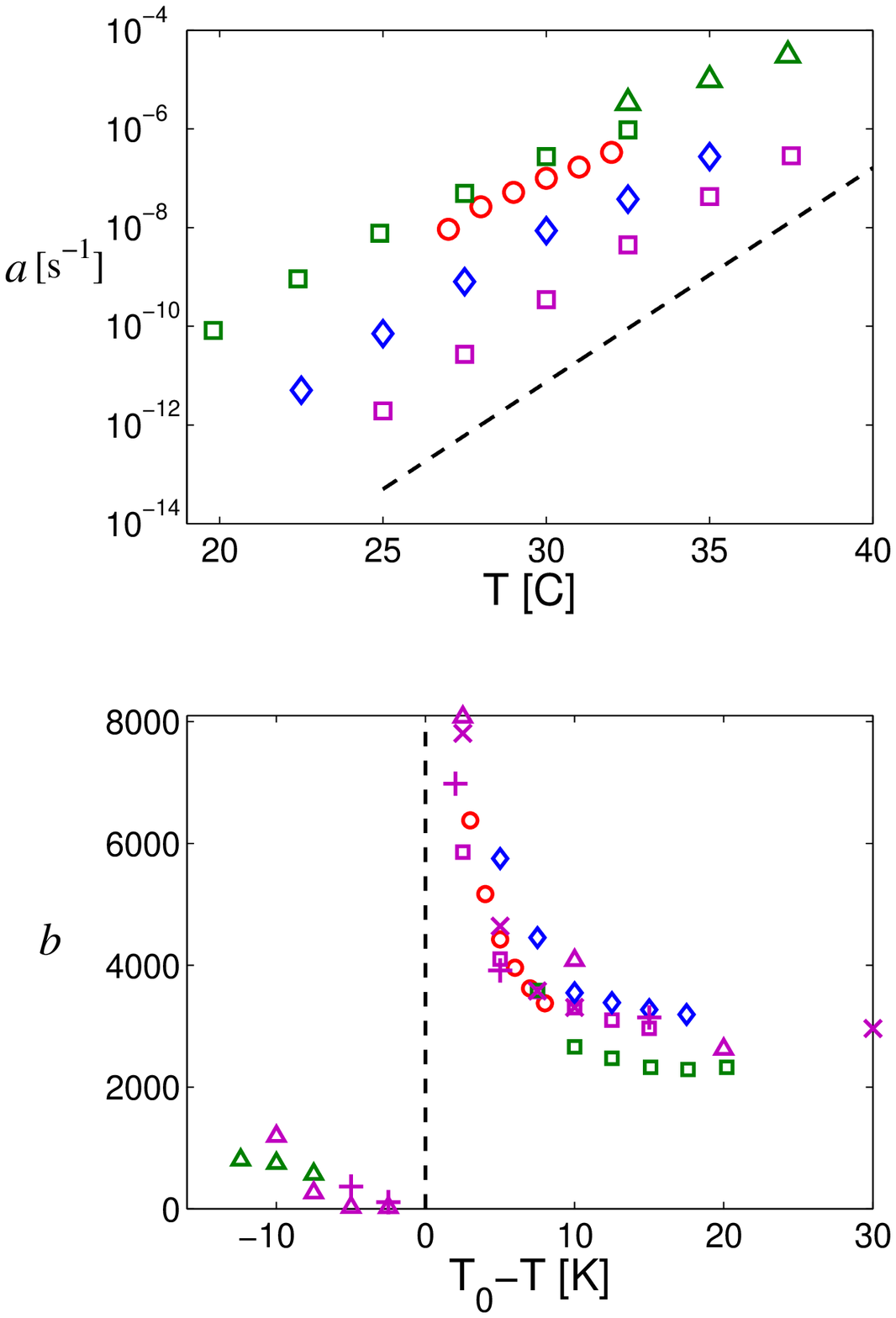}
\caption{(color online) Variation of the parameters in Eq. (\ref{model})
 with initial
temperature $T_0$ and target temperature $T$. Legend:
(\textcolor{magenta}{$\pmb{\square}$})
PVAc down-jumps from $T_0 \!=\! 40^\circ C$ \cite{Kovacs1963}
(\textcolor{magenta}{$\pmb\times$})
PVAc down-jumps to $T \!=\! 30^\circ C$ \cite{Kovacs1963}
(\textcolor{magenta}{$\pmb\triangle$})
PVAc T-jumps to $T \!=\! 40^\circ C$ \cite{Kovacs1963}
(\textcolor{magenta}{$\pmb{+}$})
PVAc T-jumps to $T \!=\! 35^\circ C$ \cite{tau_eff,Kovacs1963}
(\textcolor{red}{$\pmb\circ$})
PVAc down-jumps from $T_0 = 35^\circ C$ \cite{SvobodaPVAc}
(\textcolor{blue}{$\pmb\lozenge$})
PVAc down-jumps from $T_0 \!=\! 40^\circ C$ \cite{Kovacs1958}
(\textcolor{OliveGreen}{$\pmb{\square}$})
Glucose down-jumps from $T_0 \!=\! 40^\circ C$ \cite{Kovacs1963}
(\textcolor{OliveGreen}{$\pmb{\triangle}$})
Glucose up-jumps from $T_0 \!=\! 25^\circ C$ \cite{Kovacs1963}
(a) Variation of relaxation rate $a$ with $T$ for experiments of fixed
initial condition. The dashed line is an exponent $e^{T/\tilde{T}}$
with $\tilde{T}=1^\circ K$. The error bars are smaller than the symbols.
(b) For PVAc and glucose $b$ shows a consistent dependence on
jump magnitude $T_0 \!-\! T$.} \label{Fig3}
\end{figure}

Is it a mere curve fitting?
Equation (\ref{model}) provides an excellent description of the data, at a price of independently varying the model parameters for each
set of experimental conditions. In addition, we already know that a single variable approach as in Eq. (\ref{model}) cannot constitute a predictive model.
Nevertheless, we argue that the variation of the parameters $a$ and $b$ with experimental condition may be physically meaningful and provide us with physical insight into structural relaxation.
Indeed, Fig. \ref{Fig3} shows that $a$ and $b$ do show systematic variations with $T_0$ and $T$. Panel (a) shows $a(T)$ for PVAc and glucose under a
fixed initial temperature $T_0$ (in each experiment, both down and
up-jumps). The first observation is that $a(T)$ is a function, i.e.
it exhibits a smooth variation with $T$. The same is true for
$b(T)$, see panel (b) for measurements with fixed $T_0$.
In fact, $a$ exhibits a strong exponential dependence on $T$.
This is a manifestation of the dramatic slowing down of dynamics
associated with the glass transition. The possible origin of this
dependence in thermally activated processes will be
discussed below.

In the present context, the observation that the structural state of a glass
cannot be described by a single non-equilibrium variable implies that $a$ and $b$ should depend on the initial temperature $T_0$.
Indeed panel (b) demonstrates that $b$ depends on $T_0$.
Here the parameter $b$ in PVAc and glucose is plotted
against the jump size $T_0\!-\!T$ for both down-jumps and up-jumps
and a wide range of variation of both $T$ and $T_0$. This shows that
$b$ is in fact a function of both $T$ and $T_0$, which strikingly implies that a
relaxing glass carries memory of its original state at $T_0$ for
very long times, sometimes for days!
The dependence of $a$ on initial $T_0$ (not shown here)
is the opposite of that of $b$. In spite of this fact, we have
not found any simple connection between the two. We also verified (not
shown) that the final exponential relaxation rate, which is controlled by
the product $a(T,T_0)\,b(T,T_0)$, depends on $T_0$ as well.

Another interesting feature of panel (b) is the approximate collapse
of all measurements on a single curve as a function of $T_0\!-\!T$. This lends support to
the view that $b$ has a physical meaning. We also note the apparent
discontinuity of $b$ when passing from up-jumps to down-jumps.
Discontinuity of this kind in relaxation times is known in the
literature as the $\tau_{eff}$-paradox \cite{tau_eff}. We take it to
mean that as the asymptotic volume at $T$ is approached from above
(through a down-jump) or from below (through an up-jump), the system explores disparate
regions in phase space. This is yet another indication that
the volume alone does not tell the whole story about glassy relaxation.
This might also be related to the fact that the volume and other
thermo-mechanical properties of glasses do not
necessarily equilibrate simultaneously (see \cite{AngellReview} section B.1.8).

\begin{figure}[ht]
\includegraphics[scale=0.55]{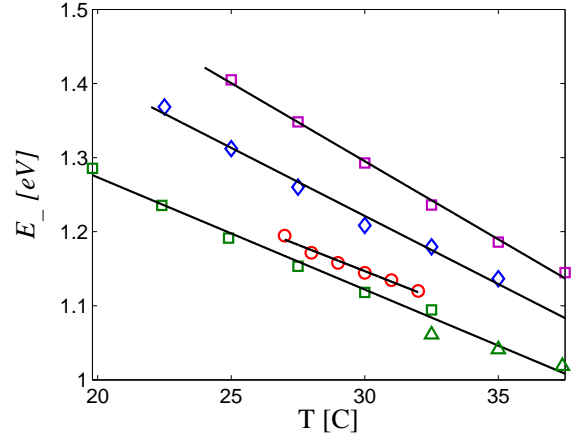}
\caption{(color online) The volume dependent
energy barrier $E_-(V_T) = -k_B T \log(a/k_0)$ for
the parameter $a$ in Fig. \ref{Fig3} panel (a). Here $k_0 = 10^{12}$ sec$^{-1}$.
Solid lines are linear fits.}\label{Fig4}
\end{figure}

Equation (\ref{model}), with its parameters $a$ and $b$, does seem
to contain meaningful physical information about the process of
structural relaxation. In order to quantify this information, we propose to rationalize the equation using a simple
two-state model. A common way to motivate such an equation
is to assume that an Arrhenius-like process depends
on an observable. In this spirit, let us assume that the volume evolves through
activated jumps of volume elements between a contracted ``$-$
state'' and an expanded ``$+$ state''. A rate equation that
describes this evolution is
\begin{equation}
\label{kinetics} \dot{n} = - k_- n +k_+ (1-n)\ ,
\end{equation}
where $n$ is the fraction of elements in the $+$ state.
A key assumption here is that the activation energies
are volume dependent, i.e.
$k_\pm \propto \exp(-E_\pm (n) /k_B T)$.
$E_\pm (n)$ are energy
barriers for expansion and contraction. They are assumed to grow
with density -- it is harder to move in a denser system. Taking into
account the fact that $\delta V/V$ is small, of the order of
$10^{-3}$, we approximate $n \!\simeq\! n_T +\delta n$, where $n_T$
is the equilibrium value of $n$ at temperature $T$. By neglecting
terms of order $\delta n$, using the equilibrium condition $k_-(n_T)
n_T\!=\!k_+(n_T)(1-n_T)$ and writing $\delta_n \!=\! \delta n /
n_T$, we find that
\begin{equation}
\label{volume} \dot{\delta}_n = k_-(n_T) \left(e^{A_+ \delta_n}
- e^{A_- \delta_n}\right) \ ,
\end{equation}
where $A_\pm\!=\! -(n_T/k_B T) \partial E_\pm(n_T)/\partial n$
are expected to be non-negative, as barriers should become smaller
with expansion. Equation (\ref{model}) then emerges as a special
case when $a(T)\!= \!k_-(n_T)$, $A_+ \!=\! 0$ and
$b(T)\!=\!A_-$.

This model allows us to extract an energy barrier scale from $a = k_0 \exp(-E_-(V_T) /k_B T)$, where $V_T$ is
the equilibrium volume at $T$. Since $k_0$ is a
pre-exponential factor, its exact value is not important. We
take $k_0\! \sim \! 10^{12}$ sec$^{-1}$, which is the typical scale of molecular vibration.
In Fig. \ref{Fig4} we extract values of
the energy barrier $E_-$ from the experiments summarized
in Fig. \ref{Fig3}(a). The activation barrier turns out to be on
a 1eV scale, which might be indicative of cooperative rearrangements of tens of monomers.

To further test the latter possibility, we estimate the size of rearranging regions by a dimensional
argument using the measured derivative $\partial E_-(V_T)/\partial T$. We define a
volume scale $v$ as
\begin{equation}
v \sim  \kappa V \pd{E_-}{V} \sim -\frac{\kappa}{\alpha}\pd{E_-}{T}\ ,
\label{volume_scale}
\end{equation}
where $\kappa\! =\! V^{-1}\partial V/\partial P$ is the isothermal compressibility
and $\alpha = V^{-1} \partial V/\partial T$ is the thermal expansion coefficient.
Linear regressions shown in Fig. \ref{Fig4} yield
$\partial E_-/\partial T \! \sim -0.02$ eV/K. Taking for PVAc
$\alpha\simeq 7 \!\times\! 10^{-4}$ K$^{-1}$ and $\kappa\! \simeq\! 0.42$ GPa$^{-1}$ \cite{Kovacs1963}
we find $v\! \sim \! 2$ nm$^3$. A similar estimate for glucose also
results in a cubic nanometer scale. These estimates imply that structural rearrangements involve tens of basic units (e.g. monomers in a polymer), which seems consistent with the typical size of dynamical heterogeneities reported in the literature \cite{Ediger}.
Therefore, while the model cannot explain the dependence of the parameters on the initial temperature $T_0$, it does seem to be sensitive to the dominant scales of the underlying relaxation processes.

In summary, we proposed a simple mean-field equation to describe
structural relaxation in glasses. Our analysis clearly demonstrates that the structural state of a relaxing
glass cannot be fully accounted for using a single non-equilibrium variable. In fact, we explicitly showed that
a glass may carry information about its thermal history for extremely long timescales.
Furthermore, the simple analysis allows the estimation of typical energy and volume scales associated
with thermally-activated structural relaxation processes. Theses estimates indicate the existence
of cooperative rearrangements on a super-molecular scale.

I.K. is very grateful to R. Svoboda and G.B. McKenna for giving
access to data and providing insightful comments. I.K. acknowledges
the support of the U.S.--Israel
Binational fund (Grant No. 2006288), the James S.
McDonnell Fund, and the European Research Council
(Grant No. 267256). E.B. acknowledges
support from the James S. McDonnell Fund, the Harold Perlman Family Foundation and the William Z.
and Eda Bess Novick Young Scientist Fund.

\end{document}